# Challenges in addressing student difficulties with quantum measurement of two-state quantum systems using a multiple-choice question sequence in online and in-person classes

Peter Hu[1], Yangqiuting Li[2], and Chandralekha Singh[1]

[1]*Department of Physics and Astronomy, University of Pittsburgh, Pittsburgh, PA 15260*
[2]*Department of Physics, Oregon State University, Corvallis, OR 97331*

**Abstract**

Research-validated multiple-choice questions comprise an easy-to-implement instructional tool that serves to scaffold student learning and formatively assess students' knowledge. We present findings from the implementation, in consecutive years, of a research-validated multiple-choice question sequence [referred to in this study as a Clicker Question Sequence (CQS)] on quantum measurement as it applies to two-state quantum systems. This study was conducted in an advanced undergraduate quantum mechanics course, in both online and in-person learning environments across three years. Student learning was assessed after traditional lecture-based instruction in relevant concepts, and their performance was compared with that on a similar assessment given after engaging with the CQS. We analyze, compare, and discuss the trends observed in the three implementations.

**Introduction**

Quantum measurement is a foundational concept in quantum mechanics (QM) which students must learn, with applications in future studies in the field and for many related fields, including the emerging field of quantum information science [1–3]. Yet because the formalism is different from that for classical measurement and difficult to ground in everyday experience, there are many concepts related to quantum measurement that can be challenging for many students to grasp [4].

Prior research suggests that students in quantum mechanics courses often struggle with many common difficulties, and research-validated learning tools can effectively help students develop a functional understanding and build a robust knowledge structure [5–30]. For example, our group has developed, validated and implemented Quantum Interactive Learning Tutorials (QuILTs) with encouraging results on many topics in quantum mechanics [4,31,32]. Other commonly used research-based learning tools in physics include clicker questions [33], which are conceptual multiple-choice questions presented to a class for students to answer anonymously, typically individually first and again after discussion with peers, and with immediate feedback.

While these questions can be successfully integrated and implemented without additional technological tools, the research presented here used an electronic response system, generally referred to as "clickers," which automatically tracked student responses in real time. When presented in sequences of validated questions, clicker questions strive to systematically help students with particular concepts that they may be struggling with. Previously, such multiple-choice question sequences, or Clicker Question Sequences (CQS), related to several key QM concepts have been developed, validated and implemented [34–38] with encouraging results. As they are effective and are relatively easy to incorporate into a typical QM course, without the need to greatly restructure other classroom activities including lectures or assignments, CQSs are a promising way to help students learn challenging concepts as a supplement to traditional lectures and homework assignments.

**Theoretical framework**

In QM courses, whose content can be difficult for students, instructors must consider research-based pedagogical approaches to engage students and help them learn these challenging, foundational concepts. Our theoretical framework for developing, validating and evaluating CQSs hinges on two different aspects of research-based pedagogical approaches, the balancing of innovation and efficiency [39] and taking advantage of peer interaction. The CQSs use research on student difficulties as resources [40] and efficiency is embedded in the way that concepts are presequenced. The CQSs also focus on providing students opportunities for innovation by means of productive struggle with new ideas through peer co-construction. While in our research presented here the concepts are first introduced through lecture-based instruction before administering the CQS, it could also be possible to structure the CQS around a just-in-time teaching scheme [41,42].

Collaborative learning can be productive in physics classrooms [33,43], particularly when individual accountability and positive interdependence have been suitably incentivized, such as through grade incentives. Peer collaboration has been shown to be an effective method for students to learn in previous work for a variety of contexts [43,44], including in physics [45]. Students' performance on conceptual physics questions can receive a substantial boost from working in pairs compared to when only working individually. In a phenomenon known as co-construction, prior research shows that student pairs in which neither student initially answered the questions correctly were able to converge on the correct answers 30% of the time, an effect that persisted when the students were assessed again individually, pointing to retention [46]. In QM courses, this rate of co-construction was about 25% [47].

Emphasizing the importance of collaborative learning, Chi et al. proposed the ICAP framework, in which there are four broad modes of learning: Interactive, constructive, active, and passive (ICAP). Only the constructive and interactive modes are concerned with the high engagement for which the CQS is designed. The main difference between the constructive and interactive modes is that, instead of providing support to students for constructing knowledge individually, the interactive mode is characterized by co-construction in small groups (realized by collaborative learning) and is shown to be associated with larger improvements [48,49].

The clicker questions, first popularized for use in physics courses by Mazur using the *Peer Instruction* technique, are intended to be conceptual multiple-choice questions posed to the class to which students reach a consensus by discussing in small collaborative groups. Mazur's method detailed in *Peer Instruction* has been associated with better learning outcomes including performance and retention [33,50], and when students engage in discussion about the CQS questions in small groups, they work under the interactive learning mode within the ICAP framework. Although this is the preferred mode, in the research presented here, constraints imposed by time and the affordances of the technology during the online implementation resulted in a largely absent groupwork component for the CQS. Instead, students were simply given the questions to think about, and they answered the questions via individual polling. Therefore, under the ICAP framework, we consider the students to be in the constructive learning mode while engaging with the CQS content in the online year, and in the interactive mode during the in-person years.

## Methods

*Development and validation*

The CQS on quantum measurement discussed here is intended for use in upper-level undergraduate QM courses. Here we summarize the development and validation of the CQS. We took inspiration from some of the previously-validated learning tools on quantum measurement to first determine the learning objectives. In particular, much research involving cognitive task analysis, from both student and expert perspectives, has already been conducted in the development and validation of a QuILT and a CQS on quantum measurement in the context of wavefunctions in an infinite-dimensional vector space [4,32]. A number of individual student interviews as well as investigations in authentic classroom environments have previously been conducted to develop, validate and evaluate classroom effectiveness of this QuILT. Using the insights from the QuILT to identify students' prior knowledge, their difficulties, and the scaffolding supports needed to help reduce those difficulties, we adapted the relevant learning objectives and questions from the QuILT while also drafting and iterating new ones for measurement related to two-state quantum systems for the CQS. This process involved the input of researchers and other faculty members, incorporating many perspectives to ensure maximal clarity and consistency in the wording and framing of the questions.

The CQS is designed to help students improve their conceptual understanding of quantum measurement, so we avoided complicated calculations to reduce cognitive overload for students. Since the CQS is designed to be used in class with peer instruction, we ensured that it was of a length suitable for administration during limited class time while still covering the common difficulties students have in understanding quantum measurement. We carefully designed each alternative choice in each multiple-choice question and incorporated the common incorrect responses that we found in previous interviews and students' written responses to the QuILT and its corresponding pretest and post-tests. Thus, the CQS provides students opportunities to think about common difficulties, struggle productively, and get immediate feedback from their peers and instructors. Another feature of the quantum measurement CQS is its inclusion of both concrete and abstract questions. Concrete questions provide opportunities for students to apply their knowledge in concrete contexts, which help them learn applications of the quantum measurement concepts in specific contexts. Concrete questions are usually followed by or integrated with abstract questions, which can help students generalize their understanding of the concepts and transfer their knowledge across contexts.

In the quantum measurement CQS, the questions are carefully sequenced to build on each other. For example, the same concept may be applied in different contexts or different concepts may be applied in similar contexts in two consecutive questions. Thus, students can compare and contrast the premise of consecutive questions to solidify their understanding of the concepts and build their knowledge structure. To facilitate class discussions after peer interaction, we developed some discussion slides between the CQS questions, which can be used by instructors to review and emphasize the important concepts in the previous questions during general class discussions on some broader themes related to those questions. These discussion slides were iterated not only amongst the researchers but also with other QM instructors.

To determine the effectiveness of the CQS, we developed and validated a pretest and post-test containing questions on topics covered in the CQS, simultaneously with and using the same process as the CQS. The post-test was a slightly modified version of the pretest, containing changes such as use of different quantum states, but otherwise maintaining underlying conceptual similarity. There are surveys that have been developed previously that ask similar questions to the ones found in the CQS, pretest, and post-test [51–53]. The pretest and post-test questions are reproduced in Appendix A.

After the initial development of the quantum measurement CQS and the pretest and post-test, starting with the learning objectives adapted from the inquiry-based guided sequences in the QuILT [29] as well as empirical data from student responses to existing individually-validated questions in previous years, we further validated them by conducting individual interviews with four students in which they completed the pretest, entire CQS, and post-test using a think-aloud protocol. In these interviews, we asked students to think aloud while answering the questions to understand their reasoning, refraining from disturbing them so as not to disrupt their thought processes. After each question, we first asked students for clarification of the points they may not have made, then we led discussions with them on each choice as appropriate. The feedback from students helped in fine-tuning and refining the new questions, as well as ensuring that they were appropriately integrated with existing ones to construct an effective sequence of questions. In the interviews, we found that students showed some common difficulties with quantum measurement consistent with results from prior studies involving the QuILT [4,32], and also that after working through the whole CQS, their difficulties with many concepts related to quantum measurement were reduced significantly. The interviewed students also reported that they found the scaffolding provided by the sequenced questions and discussion slides helpful; these slides are for instructors to discuss various issues during class discussions after students have answered a set of CQS questions.

*Learning objectives*

The learning objectives of the CQS were inspired by those for the QuILT and a CQS on quantum measurement in the context of wavefunctions [32]. CQS 1.1-1.3 help students make the distinction between quantum measurement and the action of an operator corresponding to an observable acting on a quantum state. These questions also focus on helping students be able to describe key ideas about the result of a measurement (i.e., an eigenvalue of the operator corresponding to the observable being measured) and the state of the system after a measurement (i.e., wavefunction collapse). CQS 2.1-2.3 help students calculate the probabilities of measuring certain outcomes in a given quantum state, which may or may not be written in the measurement basis. CQS 3.1-3.2 are an extension and synthesis of these ideas, in which students have to describe the results of consecutive measurements made in different bases. The last questions in the sequence, CQS 4.1-4.2, ask students to connect quantum measurement with the concept of an expectation value of an observable and its calculation (only administered to one of the classes). The clicker questions are provided in Appendix B (correct answers are provided).

*Course implementation and instructor details*

The data presented here are from administration of the validated CQS in a mandatory first-semester junior-/senior-level QM course, taught once per year during the fall semester, at a large research university in the United States. The final version of the CQS was implemented in three consecutive years, one online and two in person, with very minor adjustments made between years.

One of the two instructors who taught an in-person class was also the instructor for the online class, which enables us to draw comparisons between different classroom environments as well as different instructors.

One instructor used McIntyre's textbook *Quantum Mechanics: A Paradigms Approach*, while the other used Griffiths's *Introduction to Quantum Mechanics*. The instructor who used Griffiths's textbook started with the chapter on formalism (chapter 3), then covered spin-1/2 in chapter 4 before going back to chapters 1 and 2. Both instructors thus covered the material pertaining to two-state systems early in the course, before introducing wavefunctions and infinite-dimensional Hilbert spaces. The two instructors were very supportive of physics education research and have implemented physics education research-based pedagogies in their classes many times.

During the online implementation, the CQS was presented as a Zoom poll while the instructor displayed the questions via the "Share Screen" function. For the in-person implementations, the poll was replaced by a functionally similar classroom clicker system and students were asked to discuss their responses with each other before answering each question. For each question, the instructor displayed the results after all students had voted, before a full class discussion of the validity of the options provided. In all implementations, students were incentivized with 80% for participation and 20% for correctness on each question. Because of the difficulties in adapting to the online environment in a way that remained conducive to small-group student discussion, the Peer Instruction feature was largely forgone in the online administration in favor of mostly instructor-led discussions. Peer Instruction was realized for the in-person administrations. Therefore, referring to the ICAP framework, we consider the CQS a constructive activity in the online class and an interactive one during the in-person classes. The online class was taught after the emergency transition to remote courses, when instructors and students had some time to adapt and prepare for the norms and expectations of the remote environment, though outstanding norms and circumstances would likely have had some impact not present in more "normal" online teaching environments. That said, students in the online QM course did not reveal to the course instructor any additional challenges in pursuing their studies remotely, suggesting that they were still able to participate in the course reasonably well.

In both online and in-person classes, students first learned from traditional lecture-based instruction all the concepts covered in the CQS before completing the pretest. After administration of the CQS over two to three class sessions, students completed the post-test. During this time period, the instructors did not cover the material again, except during the class discussions afforded by the CQS, but students were given traditional homework sets from the course textbooks. However, in our prior research, student post-test performance is significantly improved when research-based tools are used in instruction, as compared to control groups of students who only have traditional instruction and homework [32,54]. That said, this study is quasi-experimental [55] in design, in light of these and other factors over which we did not have complete control.

As an additional point, the CQS as designed is not intended to serve as students' first instructional experience in the quantum measurement concepts covered. In our implementations, students received lecture-based instruction to give them the preparation to engage on a deep level with the questions. However, it is not imperative that students are prepared by lecture-based instruction. As an example, just-in-time teaching could also accomplish this [41], in which students are assigned videos or readings and given some assessment tasks based upon them before class, though we did not investigate this possibility in the present study. Thus, efficiency is embedded in this combination of lecture and CQS instruction as the CQS questions build on each other. At the same time, instructors foster innovation by providing students opportunities to productively struggle with challenging concepts via peer interaction and full class discussions. The student-to-student interactions may also allow students opportunities for co-construction of knowledge.

Students knew in advance that they would take the pretest after traditional lecture-based instruction on relevant concepts (but before the CQS) and the post-test after the CQS. Both pretests and post-tests were framed as quizzes, but students were graded on the pretest for completion and post-test for correctness. The in-class rubric for the post-test grading was more lenient than the one used for this research, so the assessments were not particularly stressful. The quizzes were closed-book and closed-notes. All classes were given roughly half a class period (about 25 min) to take the quizzes synchronously, but all students in the online group exercised the option to keep their cameras turned off. Most of the questions on the pretest and post-test were graded on an all-or-nothing scheme. We would have liked to give students opportunities to provide their reasoning, but due to concern over time constraints in administering the pretests and post-tests in the class, in addition to our work being qualitatively supported by student reasoning from numerous prior studies [4,16,51,53,56–58], our questions took the form presented in Appendix A. Question 3 was graded on a three-tiered scale (zero, half, and full credit), for which two researchers graded half of the pretest and post-tests and, after discussion, converged on a rubric for which the inter-rater reliability was greater than 95%. Afterward, one researcher graded the remaining half of the tests. A detailed breakdown of student performance on the tested concepts is provided in the next section. In the closing sections, we compare the online implementation with both an in-person implementation with the same instructor as well as one with a different instructor. We also compare the two in-person implementations with each other to determine the generalizability of the CQS's usefulness.

**Results and Discussion**

In this section, we present results of implementation of the final version of the CQS in three consecutive years. The concepts that the questions dealt with are summarized in Table I. Results for each question, as well as normalized gains [59] and effect sizes [60], are listed by class in Table II, with modality and sample size specified for each case. These data are represented visually in bar charts in Figure 1.

**Table I.** Summary of the concepts that were covered in the CQS, listed along with the pretest/post-test questions and CQS questions that address them.

| Concept | Pre-/post-test question | Corresponding CQS questions |
|---|---|---|
| $\hat{S}_z\|\chi\rangle = \frac{\hbar}{2}\|+z\rangle$ or $\hat{S}_z\|\chi\rangle = \frac{\hbar}{2}\|-z\rangle$ represents measurement of z-component of spin, where $\|\chi\rangle$ is the state before the measurement and $\|\pm z\rangle$ is the state after the measurement [incorrect statement] | 1 | 1.1 |
| Same as question 1, but with general Hermitian operator $\hat{Q}$ | 2a | 1.2, 1.3 |
| Normalization of a quantum state | 2b | 1.3, 2.1 |

| Measurement in a quantum state that is not given in the measurement basis | 2c | 2.2, 3.2 |
|---|---|---|
| Possible outcomes, and probabilities of measuring those outcomes, for a measurement of an observable in a given quantum state | 3 | 1.3, 2.1 |
| Normalized state after measurement collapse | 4a, 4b | 2.3, 3.1, 3.2 |
| Successive measurements of $S_x \to S_z$ | 4c | 3.1, 3.2 |
| Successive measurements of $S_x \to S_z \to S_x$ | 4d | 3.2 |
| Expectation values of various observables in various states | 5 | 4.1, 4.2 |

**Table II.** Results of the online and in-person administrations of the CQS. Comparison of pretest and post-test scores, along with normalized gains [59] and effect sizes as measured by Cohen's $d$ [60], for students who engaged with the CQS. For the online class, $N = 30$; for in-person class 1, $N = 23$; and for in-person class 2, $N = 28$ (note: an additional question 5 was added for in-person class 2).

| Online class | Question # | Pretest | Post-test | Normalized gain | Effect size |
|---|---|---|---|---|---|
| | 1 | 27% | 63% | 0.50 | 0.78 |
| | 2a | 33% | 63% | 0.45 | 0.62 |
| | 2b | 97% | 97% | - | - |
| | 2c | 67% | 83% | 0.50 | 0.39 |
| | 3 | 92% | 97% | 0.60 | 0.24 |
| | 4a | 73% | 87% | 0.50 | 0.33 |
| | 4b | 80% | 97% | 0.83 | 0.53 |
| | 4c | 77% | 93% | 0.71 | 0.47 |
| | 4d | 67% | 93% | 0.80 | 0.70 |
| In-person class 1 | Question # | Pretest | Post-test | Normalized gain | Effect size |
| | 1 | 26% | 65% | 0.53 | 0.84 |
| | 2a | 26% | 61% | 0.47 | 0.73 |
| | 2b | 100% | 96% | - | - |
| | 2c | 87% | 91% | 0.33 | 0.14 |
| | 3 | 83% | 96% | 0.75 | 0.58 |
| | 4a | 52% | 78% | 0.55 | 0.56 |
| | 4b | 65% | 91% | 0.75 | 0.65 |
| | 4c | 70% | 70% | - | - |
| | 4d | 43% | 57% | 0.23 | 0.26 |
| In-person class 2 | Question # | Pretest | Post-test | Normalized gain | Effect size |
| | 1 | 36% | 46% | 0.17 | 0.22 |
| | 2a | 21% | 68% | 0.59 | 1.04 |
| | 2b | 100% | 96% | - | - |
| | 2c | 79% | 86% | 0.33 | 0.18 |
| | 3 | 86% | 89% | 0.25 | 0.14 |
| | 4a | 54% | 71% | 0.38 | 0.37 |
| | 4b | 68% | 75% | 0.22 | 0.16 |
| | 4c | 46% | 68% | 0.40 | 0.44 |
| | 4d | 46% | 50% | 0.07 | 0.07 |
| | 5 | 30% | 73% | 0.62 | 1.00 |

**Figure 1.** Bar charts representing the pretest and post-test scores for all classes, as reported in Table II. Error bars represent standard error.

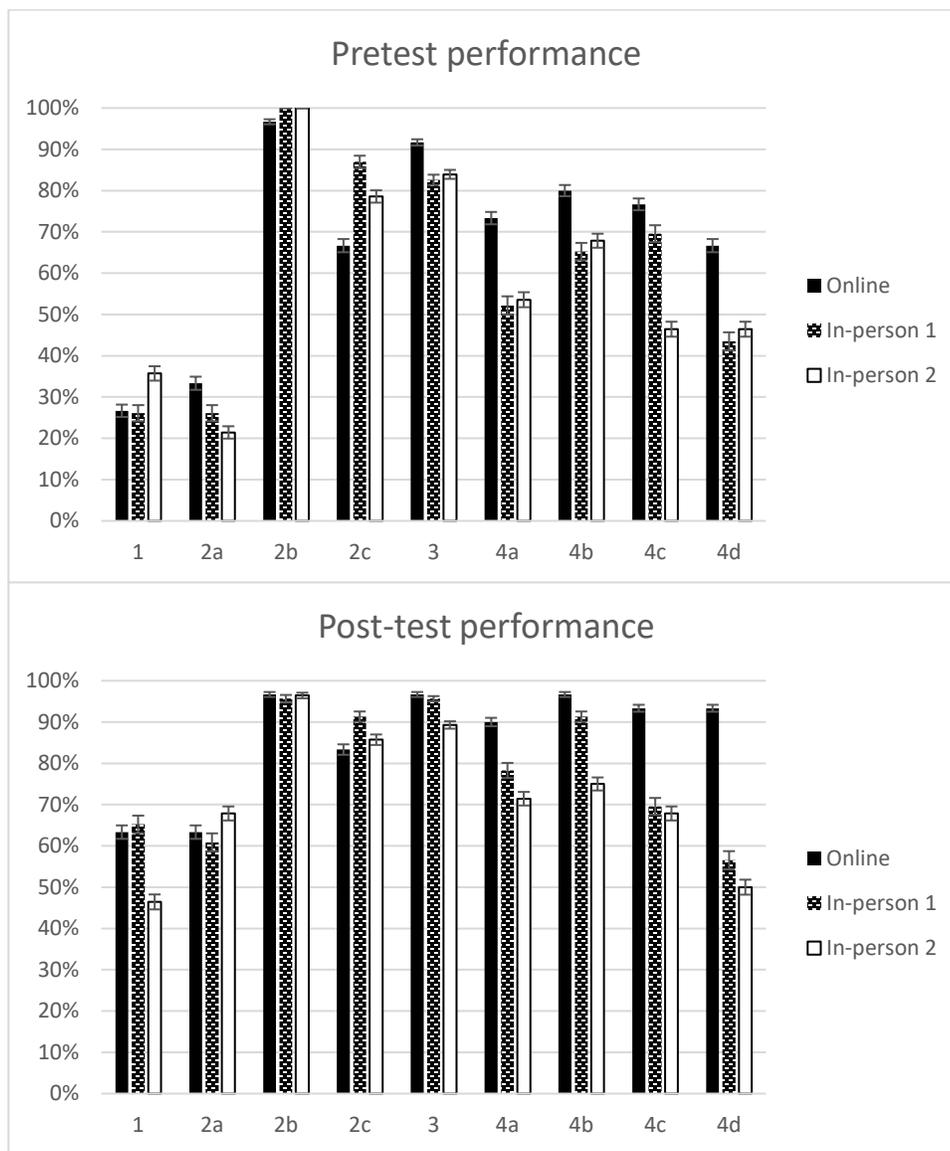

Overall, the results are encouraging, and students performed well on the post-test, with reasonably high normalized gains. Effect sizes were generally medium to large for the online class and in-person class 1, and overall they were more modest for in-person class 2.

These results in Table II suggest that the concepts assessed by questions 2b and 3 may be understood well after traditional lecture, while the concept assessed by 1 and 2a is much more difficult for students. Although improvement is seen on student performance on questions 1 and 2a after the CQS (see Table II), the underlying concept still appears to be difficult. Later we will discuss that, after struggling on both the pretest and post-test, students improved further on corresponding later midterm exam questions, after being provided solutions to the post-test questions (see subsection titled "*Retention and further learning after post-test solutions were made*

*available*"). For questions 4a-d, the improvement is also noticeable in all classes for a great majority of the questions.

Given that the two in-person years were taught by different instructors, and that there are improvements in student performance for both years, it is clear that the CQS is beneficial for students despite the differences between instructors' approaches. Furthermore, the CQS is beneficial in both online and in-person environments. As this study was quasi-experimental, it is possible that there are other effects in addition to the CQS itself that led to these benefits. However, in a prior study, a control group of students, who were given the post-test on quantum mechanics concepts after traditional lecture-based instruction and associated homework, were significantly outperformed ($p < 0.0001$) by three sets of experimental groups of students who engaged with clicker questions after their traditional lectures and had homework similar to the control group [54]. Moreover, in another study [32], a control group of students who only had traditional lecture and homework were outperformed by students who used research-based tools after traditional lecture and homework. Thus, we believe that the CQS plays an important role in these improvements, and that homework alone does not provide the benefits seen here. Furthermore, a prior study in introductory mechanics comparing lecture vs microcomputer-based labs suggests that a research-based tool is a significant contributor to better performance, and that simply repeating or expanding upon previously covered material, such as through traditional problem-solving exercises, does not necessarily yield the same positive results [61].

In the following sections and Table III, we categorize student difficulties observed during the administration of the CQS, and the extent to which they were successfully addressed.

*Action of an operator corresponding to an observable on a quantum state being confused with a measurement*

Broadly speaking, when given an observable $Q$ with corresponding Hermitian operator $\hat{Q}$, eigenvalues $q_1$ and $q_2$ and eigenstates $|q_1\rangle$ and $|q_2\rangle$, many students state that a measurement of observable $Q$ is represented mathematically by $\hat{Q}|\chi\rangle = q_1|q_1\rangle$ or $\hat{Q}|\chi\rangle = q_2|q_2\rangle$ [4]. The claim is that the state before the measurement is $|\chi\rangle$, and after the measurement it is $|q_1\rangle$ or $|q_2\rangle$ depending on the result of the measurement, despite the fact that neither equation is mathematically valid. In all three implementations, students tended to gravitate strongly to this idea on the pretest (questions 1 and 2a) after lecture-based instruction and the first CQS question in this sequence (CQS 1.1). The performance on CQS question 1.1 gave the instructor an opportunity to discuss with students and deconstruct both the correct and incorrect ideas involved, such that students were able to answer correctly at a much higher rate for the following questions CQS 1.2-1.3. This improvement was reflected on the post-test, but a significant fraction of students (around 40%) continued to mistakenly rate a statement such as $\hat{Q}|\chi\rangle = q_1|q_1\rangle$ or $\hat{Q}|\chi\rangle = q_2|q_2\rangle$ as true (see Table II). It appears that this difficulty is so persistent that students underperform on this question compared to the rest of the post-test. However, as we will discuss later, the improvement is encouraging on another assessment given two weeks after the post-test (discussed in the subsection "*Retention and further learning after post-test solutions were made available*").

Prior studies [4,51,56,52],have shown that when the operator in question is the Hamiltonian, students struggle with this concept. In a multi-institutional study, only 22% of undergraduate and 26% of graduate students managed to correctly answer a question on the Quantum Mechanics Formalism and Postulate Survey (QMFPS) targeting this concept [53]. It is possible that students are improperly overgeneralizing the time-independent Schrödinger equation (TISE) in some ways, rather than restricting it to only the cases where the eigenvalue equation holds. We also find analogous patterns in student responses when asked about a generic operator. This may also indicate an incomplete knowledge of linear algebra, with students not knowing that an operator is a linear transformation acting on the ket state, and so dropping one of the eigenstates from the right-hand side would not serve as a prominent red flag. In fact, in interviews in which students agreed with $\hat{Q}|\chi\rangle = q_1|q_1\rangle$ or $\hat{Q}|\chi\rangle = q_2|q_2\rangle$, when it was pointed out that this type of equality violates the rules of linear algebra, only then did some notice the issue, while others suggested that quantum mechanics itself might simply not follow linear algebra [16]. These students thought that quantum measurement must be represented by some equation that resembles the TISE. The overgeneralization takes this particular form likely because of the strong emphasis that a measurement of an observable in any state can only yield the eigenvalues of the operator corresponding to the observable, and that the state thereafter has collapsed to the eigenstate associated with the eigenvalue that was obtained. Coincidentally, the TISE contains entities that represent the operator, the eigenvalues, and the eigenstates of the operator, even though the TISE (and equations involving operators in general) have nothing to do with the measurement process. Ultimately, there are many surface features that tie the two ideas together, and significant care and persistence, e.g., via clicker questions and other research-based tools, is needed to disentangle them from each other.

*Normalization of a quantum state*

Pretest and post-test question 2b concerned the normalization of expansion coefficients of a quantum state written in a particular basis, and students did very well on both the pretest and post-test (see Table II). The normalization of a quantum state after state collapse following a measurement, which proved to be a challenging concept, was assessed by question 4a. The question started with a given state and a measurement of $S_x$, yielding a particular eigenvalue (e.g., $-\frac{\hbar}{2}$, which indicates a collapse into the state $|-x\rangle$). Some students provided the original state as their state after measurement rather than the proper eigenstate of $\hat{S}_x$. Other incorrect answers included keeping the expansion coefficient of this eigenstate without normalizing (e.g., $\sqrt{\frac{3}{10}}|-x\rangle$) or by attaching the eigenvalue itself to the state as the result of the measurement (i.e., $-\frac{\hbar}{2}|-x\rangle$), a response consistent with the difficulty discussed in the preceding section. Such answers were overall observed less frequently on the post-test in all three classes, indicating the effectiveness of the CQS in helping students with this concept.

Normalization of a quantum state is not necessarily an intuitive thing that first-time learners check for at every step of a calculation, so keeping coefficients like $\sqrt{\frac{3}{10}}$ does not seem unnatural [52,57]. Indeed, in a multi-institutional study, when asked on the QMFPS [53], 17% of undergraduate students and 23% of graduate students selected such a response. Other responses like $\pm\frac{\hbar}{2}|\pm x\rangle$, or attachment of the appropriate eigenvalue of any quantum state, are likely closely related to the difficulty described above conflating the TISE, or any equation involving any operator that corresponds to an observable, with the process of quantum measurement.

*Measurements made in a basis different from the one given*

Question 2c states that a measurement of $S_y$ in a state $a|+z\rangle + b|-z\rangle$ would yield an outcome of $\frac{\hbar}{2}$ with probability $|a|^2$. Students were asked whether they agreed with this statement, which is false because it is necessary to change to the appropriate measurement basis before interpreting the meaning of the expansion coefficients. In the problem statement, specific attention was drawn to the fact that the measurement was not of the observable $S_z$, and students then should know that a basis change is necessary. Students in the in-person years did reasonably well on the pretest and also improved somewhat on the post-test, approaching full scores, i.e., everyone having relevant knowledge based upon their performance (see Table II). In a multi-institutional study, 39% of undergraduate students and 28% of graduate students did not first change the basis on a QMFPS question of this type, and this appeared to be reflected in the online class, though the necessity of a basis change was not made as salient in the question in the QMFPS [53] as in our pretest and post-test questions.

In the case of measuring components of spin for two-state systems in two-dimensional Hilbert spaces, the eigenvalues are the same regardless of which component of spin is measured. This can make things less cognitively demanding, but also may pose some difficulties when generalizing to infinite-dimensional Hilbert spaces, such as with the observables position and momentum. Transforming from one basis to another, for instance, is more mathematically complex for wavefunctions in infinite-dimensional Hilbert space even though the underlying concept is the same, and students have been found to struggle with successive measurements of energy and position for wavefunctions [57], which requires changing the basis.

*Outcome and probabilities of a measurement outcome*

In Question 3 on the pretest and post-test, students were provided a quantum state and asked about the possible outcomes of measurement, and the probabilities of obtaining those outcomes. As with question 2b, this was also a relatively easy question for students even before engaging with the CQS (see Table II). The concepts covered by these two questions are related in that normalization of a quantum state directly makes use of the fact that all probabilities must sum to 1, indicating that students had a good understanding of how to interpret probabilities of measurement outcomes in a normalized quantum state after lecture instruction. They still exhibited some improvement after CQS administration.

Students were given half credit for providing the possible outcomes, and half credit for providing the respective probabilities. The distributions of scores on this question are shown in Figure 2.

**Figure 2.** A bar chart showing the shifts in distributions for question 3 from the pretest to the post-test for each class. (A table is also provided.)

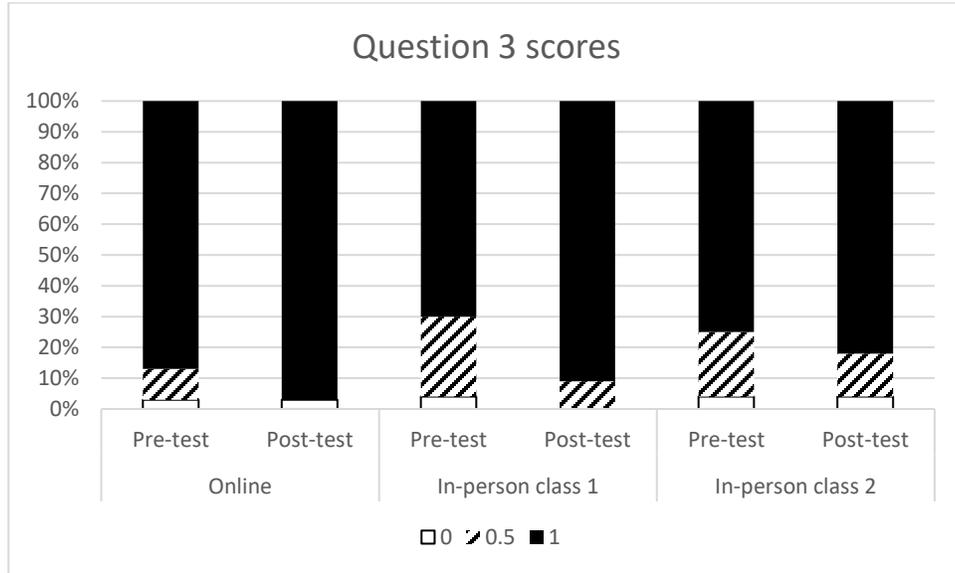

| Question 3 scores | | 0 | 0.5 | 1 |
|---|---|---|---|---|
| Online | Pretest | 3% | 10% | 87% |
| | Post-test | 3% | 0% | 97% |
| In-person class 1 | Pretest | 4% | 26% | 70% |
| | Post-test | 0% | 9% | 91% |
| In-person class 2 | Pretest | 4% | 21% | 75% |
| | Post-test | 4% | 14% | 82% |

*Results of consecutive measurements of spin components*

Questions 4b-4d asked students about the results of consecutive measurements of different components of spin in immediate succession. Pretest performance is seen to be quite high in the online administration, and less so in the in-person administrations (see Table II). Different questions had different normalized gains and effect sizes over the three years, with some questions showing little improvement in one year but large improvements in the remaining two.

Question 4b stipulated that a measurement of $S_x$ in a given state resulted in a particular eigenvalue, and asked students about the probability of obtaining the other eigenvalue from a measurement in immediate succession. Most students answered this question correctly on the post-test, although in-person class 2 had a higher error rate than the others. However, there was no discernible underlying pattern to students' incorrect answers.

Question 4c asked students about the results of a measurement of $S_z$ in the given state after the preceding measurement of $S_x$ (from question 4b) had been made. Most commonly, students who struggled changed the *original* state to the *z*-basis for their answers, not realizing that the state had collapsed to an eigenstate of $\hat{S}_x$ from the preceding measurement. This is a rather complicated question that requires students to utilize several quantum measurement concepts at each step, and it is likely that they did not recognize some parts of the question that implied the measurement collapse of a state, even if they are able to recognize the state collapse when asked directly about it. Where students were observed to improve on the post-test, they did so by rather large margins, though no improvement was observed in the first in-person class.

Question 4d provided a situation of far transfer from the CQS involving three successive measurements of incompatible spin components, e.g., $S_x \to S_z \to S_x$. Performance on this question, which has no direct analogue in the CQS, was generally lower than for many other questions, which is not surprising. That said, students who provided correct answers consistently pointed to how each measurement of a particular component of spin "destroys" knowledge of the other spin components (which was an idea emphasized in the CQS), so that the two measurements of $S_x$ will not necessarily match each other. Very few students on the post-test answered that the two measurements of $S_x$ should match in this situation.

For all parts of question 4, a common incorrect response involved using the original state (not the collapsed state following the measurement of $S_z$) to respond to the questions. This was seen on both the pretest and the post-test, somewhat less frequently on the latter. It is possible that such students either did not notice that the state should have already collapsed, or did not think such information was relevant. Prior research shows that some students think that a quantum state will return to its original state after enough time (even if it collapsed into another state); that measurement does not affect the state (which is only true for the eigenstate of the operator corresponding to the observable measured); or that discrete observables can change drastically between measurements (as opposed to continuous variables); these ideas could have also played a role [51,56,57]. Additionally, for questions 4c and 4d, some students who struggled appeared to, for instance, use the given state $|\pm z\rangle$ to incorrectly conclude that the probability of measuring a particular value for $S_x$ is zero, not realizing that $|\pm z\rangle = \frac{1}{\sqrt{2}}(|+x\rangle \pm |-x\rangle)$.

Question 4 as a whole was designed to probe students' knowledge on the collapse of quantum states when specific spin components are measured. Each successive part goes deeper into a hypothetical situation in which observables are measured whose corresponding operators either do or do not commute. Students who, for instance, provide the correct state after the first measurement of $S_x$ may not recognize that this is the state in which the next measurement must be made, or that after an intervening measurement of $S_z$ is made, this system may not yield the same outcome if $S_x$ is then measured again. It is illustrative to consider students' success and improvement in answering all four parts of the question. This is shown in Figure 3. It is clear that, on the post-test, students who answered three or four of the parts correctly increased in number, reaching 50% of the class or higher, while the number of students who correctly answered only two, one, or none of the parts decreased.

**Figure 3.** A bar chart showing the shifts in distributions for all parts of question 4 from the pretest to the post-test for each class. (A table is also provided.)

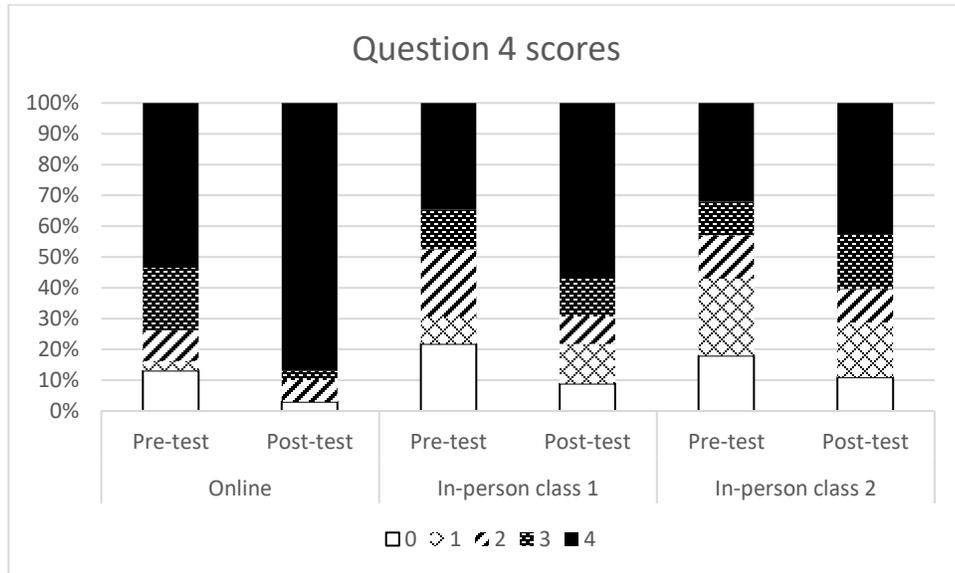

| Question 4 scores | | 0 | 1 | 2 | 3 | 4 |
|---|---|---|---|---|---|---|
| Online | Pretest | 13% | 3% | 10% | 20% | 53% |
| | Post-test | 3% | 0% | 7% | 3% | 87% |
| In-person class 1 | Pretest | 22% | 9% | 22% | 13% | 35% |
| | Post-test | 9% | 13% | 9% | 13% | 57% |
| In-person class 2 | Pretest | 18% | 25% | 14% | 11% | 32% |
| | Post-test | 11% | 18% | 11% | 18% | 43% |

*Preliminary investigation of difficulties with expectation values*

Question 5 was added to the second in-person class to investigate students' understanding of the expectation value of an observable, which prior research [58] has suggested is a challenging concept. In this administration during the second in-person class, students were additionally asked to calculate the expectation value of $S_x$ in a given state. A wide range of responses appeared on the pretest; overall, the inconsistency in these responses, as well as the number of students who left the question blank or answered "I don't know," indicates that students were not confident in their knowledge of expectation values.

The following responses on the pretest were present on the post-test in highly reduced numbers. One pattern of response was to list the eigenvalues that could result from a measurement made in the given state, sometimes alongside the probabilities with which they could be measured. Other students responded by correctly observing that any measurement made in the same initial state would yield results with the same probability distribution, but without showing any attempt at calculating the expectation value itself. Students in both groups were given no credit, since a different question had already asked them to provide the outcomes and probabilities of measuring each outcome in a given state. This is reminiscent of students' answers to a question, in a separate study, about measurement uncertainty. When asked to calculate a nonzero uncertainty in the measurement of a particular observable, some students appealed to the fact that the given state had a nonzero chance of yielding either outcome, but did not proceed to calculate a numerical value for the uncertainty [37]. Students providing such responses when asked to calculate the expectation value may not know or remember what an expectation value is, and are only able to produce part of the relevant knowledge. On the pretests and post-tests, some responses to the expectation value question had phrases such as "expected value," seeming to interpret the question as another way of asking for the possible measurement outcomes (i.e., "what values would one expect from a measurement of this observable?"). These are some possible explanations for the types of responses that students provided with some frequency that identified the outcomes and the respective probabilities of measuring each one.

Some students on the pretest, and nearly all students on the post-test, chose a valid method to write or calculate the expectation value. The dominant methods were (i) $\left(\frac{\hbar}{2}\right) P\left(\frac{\hbar}{2}\right) + \left(-\frac{\hbar}{2}\right) P\left(-\frac{\hbar}{2}\right)$, where $P\left(\pm\frac{\hbar}{2}\right)$ represents the probability of measuring $\pm\frac{\hbar}{2}$, and (ii) $\langle\chi|\hat{Q}|\chi\rangle$. Despite this, some students were unable to successfully complete the calculation. Most commonly, this was a result of choosing to calculate the expectation value via method (i), but simply finding incorrect probabilities of each measurement outcome. Students also chose method (ii) but were left confused on how to proceed. Such students received partial credit. These difficulties mirror some of those found in previous studies [4,56,58], and the multi-institutional QMFPS study found that 54% of undergraduate students and 57% of graduate students similarly struggled with finding a correct expression for expectation value [53].

All these difficulties were observed in much smaller numbers on the post-test. Furthermore, the fact that many of the difficulties found in previous work, e.g., Ref. [58], appear to have been avoided could be a sign that the CQS is helpful in guiding many students toward a productive conception of expectation value, or at least a fluency in calculating expectation values that could be useful in further consolidation of knowledge with continued instruction. Though the data are from a single year of administration of CQS, we find that students may need help beyond traditional lecture-based instruction in achieving facility with expectation value for two-state systems. Furthermore, we are optimistic that the CQS is beneficial as a vehicle for enabling students to acquire such facility.

A summary of the difficulties found, along with the CQS number and the extent to which the average student performance improved from the pretest to post-test, is shown in Table III.

**Table III.** Student difficulties addressed by the CQS questions, which are found in Appendix A. For each difficulty, previous studies are cited which have reported such observations.

| Difficulties | CQS # | Pre-/post-test # | comments |
|---|---|---|
| Action of an operator corresponding to an observable on a state confused with measurement [4,16,51,52,56] | 1.1, 1.2, 1.3 | 1, 2a | Major improvement, but still difficult |
| Normalization of a quantum state [52,57] | 1.3, 2.1, 2.3, 3.1, 3.2 | 2b, 4a | Students performed well on normalization questions in isolation (high correctness), but are less adept at providing a normalized collapsed state |

| | | |
|---|---|---|
| | | after a measurement (medium improvement) |
| Measurement basis is different from given basis [57] | 2.2, 3.2 | 2c | Some improvement |
| Outcomes and probabilities of measurement outcomes for an observable, measurement basis matches given basis | 1.3, 2.1 | 3 | High pretest and post-test correctness |
| Measurement of various spin components in immediate succession [47,51,56,57] | 3.1, 3.2 | 4b-4d | Some improvement |
| Difficulties with expectation values [4,56,58] (preliminary results) | 4.1, 4.2 | 5 | Major improvement |

*Comparisons between instructors and learning environments*

The two in-person classes with different instructors have quite similar profiles on pretest and post-test performance. Final scores on the post-test match closely when examined question-by-question, with the largest discrepancies found in questions 1 and 4b (see Table II). It is unclear why, for the second in-person year, performance on questions 1 and 2a is so different, as both are intended to target similar concepts, but it is possible that the additional explanation that question 2b provides in interpreting the action of an operator on a quantum state was enough to cue these students into rejecting the statement, while accepting the statement provided in question 1 without giving it much further thought. For question 4b, the most common incorrect answers were discussed earlier in the subsection "*Normalization of a quantum state*," and most of the students who provided such answers were in in-person class 2. Other moderate differences between the in-person years may be due to differences in instructors' approaches to the material, or the way students understood the material or the questions, but as a whole, there are not many such differences. That the instructors used different textbooks that are usually tuned to different approaches to teaching QM (although both instructors covered two-state systems early in the course) also suggests that the CQS is flexible enough to remain robust in a variety of QM curricula backed by different course textbooks.

On the other hand, comparing the online class with in-person class 2, which is the most robust comparison of the two environments because the instructor was the same, it may appear surprising that students performed largely better on the post-test in the online class despite not having access to collaborative learning opportunities. However, research [62,63] suggests that student behavior can differ depending on many factors, including stakes, between online and in-person assessments. In light of the consistency across the two in-person classes, it may be the case that the online scores are inflated.

In particular, we acknowledge that some students may have been able to consult resources that they were not intended to access during the online-administered pretest and post-tests. Even though, much like a closed-book and closed-notes quiz in an in-person class setting, the online pretests and post-tests were administered synchronously with students submitting their work at the end of the allotted time, the situations are still not quite the same, especially when all students had their cameras off. There was no way to adequately determine or enforce which resources they used during the examination time. As such, the results from the in-person classes may be regarded as more representative of performance without consulting any resources, with the pretests and post-tests administered in a more controlled environment. Another explanation, not mutually exclusive, is that the increased availability of class materials such as recorded lectures during online classes could have given students a wealth of material to study from in preparation for the pretest and post-test.

Regarding the emergency remote teaching arrangements of the online class, while data do not exist for retention of concepts in the same capacity as in the second in-person class (discussed in the "*Retention and further learning after solutions were made available*" section that follows), the final overall grade distributions were comparable between the online class and in-person class 2. Though not a perfect measure, this offers some assurance that the two classes were not extremely different in the overall level of student knowledge and learning of QM. These findings suggest that the CQS offers meaningful benefits for all the classes in which it was implemented, even with their very different course structures and social backdrops.

*Retention and further learning after post-test solutions were made available*

Students were provided neither the CQS questions nor the pretest questions as resources before administration of the post-test; however, after students were graded on their post-test performance, the post-test solutions were posted on the course learning management system. We find that students performed very well on similar questions that were asked two weeks later on the midterm examination of the second in-person class. The post-test questions that had very similar analogues on the midterm exam were 2a, 2b, 2c, 3, 4a, and 4b. On these questions, correctness was nearly 100% for each, a heartening sign that almost all the students are getting what they should out of the overall instruction after having struggled productively on the pretest, CQS, and post-test. One question merits special attention. Question 2c, which assesses whether students would be able to deduce the need to change the basis before providing the probability of measuring a particular outcome, focuses on a concept that appeared on the post-test as both a nearly exact reproduction and as a prerequisite for another question. In isolation, when explicitly asked to change the basis, students were able to do so at a nearly perfect rate, 93%, but when implicitly required as one step of another problem related to measurement, 76% of students recognized that a change of the basis was necessary to solve the problem. This suggests that, while students may understand that they must apply this concept when it is the primary focus of a question, they may not yet recognize its applicability when this concept is applicable but the question does not explicitly ask for it. For example, they may not check whether the state is written in the appropriate basis or transform to the appropriate basis when asked about outcomes of measurement, particularly when other features are present (see Table IV).

We note that prior research in introductory physics and quantum mechanics suggests that students who self-diagnose and evaluate their mistakes on earlier problem-solving tasks (e.g., quizzes or exams) are likely to do significantly better on those concepts on future exams, but that the act of providing solutions to students alone does not trigger such self-diagnostic behavior. This is true even if the students know that the material could show up on future testing, and the questions asked are identical [64,65]. Without explicit grade incentives and encouragement, many students tend not to learn from their mistakes by comparing their work to that presented in the provided solutions. Indeed, it was only when students were given an explicit reason to do this (e.g., being given a grade incentive for correcting their mistakes) that their performance on the second assessment improved.

We hypothesize that the time dedicated to struggling with the concepts may also be a powerful way to enable students to productively engage with the relevant material ahead of the exam. It is possible that experience with productive struggle in collaborative learning during the CQS activity cued students into a learning mode, making them attuned to their mistakes, and motivated to clear up their difficulties. As a complementary hypothesis, students were in a position to better understand and make use of their available resources (which included post-test solutions) after engaging with the CQS than they would have been without it. In particular, students appear to have retained or learned the measurement concepts well between their post-test and midterm exam weeks after they had gone over the relevant material during class.

**Table IV.** Student performance on similar questions given on a midterm exam about two weeks after the post-test, for in-person class 2 ($N = 28$). The concept covered by question 2c appeared in two separate questions, each in a different context.

| Question # | Post-test | Midterm | Normalized gain | Effect size |
|---|---|---|---|---|
| 2a | 68% | 93% | 0.78 | 0.65 |
| 2b | 96% | 100% | 1.00 | 0.27 |
| 2c | 86% | 93% | 0.50 | 0.23 |
|    |     | 76% | - | - |
| 3 | 89% | 96% | 0.67 | 0.32 |
| 4a | 71% | 79% | 0.25 | 0.16 |
| 4b | 75% | 100% | 1.00 | 0.80 |

**Conclusions**

Validated clicker question sequences can be effective tools when integrated with classroom lectures. We developed, validated, and found encouraging results from implementation of a CQS on the topic of quantum measurement for two-state systems, in both online and in-person settings. We drew two comparisons among the implementations, one between two different instructors for identical modes of instruction (in-person), and one between different modes of instruction (online vs. in-person) for the same instructor. Post-test scores improved for nearly every question in each implementation, with exceptions typically being questions that had very high pretest performance to begin with. Our student interviews here and in past work [4,16,51,53,56–58] have indicated that students have many conceptual difficulties with the measurement concepts examined here. By building on this work to specifically address these difficulties, the evidence of improvement after CQS instruction is likely to be due to learning rather than simply memorizing the correct answers.

In summary, it appears that the CQS provided noticeable and meaningful benefits in all observed cases despite differences in modality and instructors' lecturing styles. Within the ICAP framework, both interactive (in-person classes with peer interaction) and constructive (online classes without peer interaction) modes appear to be beneficial for students. In particular, it greatly reduces widespread student difficulties, e.g., confusing the action of an operator corresponding to an observable on a state with the act of quantum measurement. For in-person class 2, the data from a midterm exam given later in the semester indicate that students further strengthened their understanding of these concepts in the intervening time period when solutions to the post-test were provided, likely owing to the high levels of engagement and productive struggle that they experienced during the collaborative learning and class discussions of the CQS. Clearly, though, modality and instructor have some effects, whose nature will be investigated in future studies. In particular, while the CQS was useful in all of these cases, it is possible that these factors along with the choice of textbook could explain differences not only in the post-test but also in the pretest.

**Acknowledgments**

We thank the NSF for Grants No. PHY-1806691 and No. PHY-2309260. APC charges for this article were fully paid by the University Library System, University of Pittsburgh. We thank all students whose data were analyzed and Dr. Robert P. Devaty for his constructive feedback on the manuscript.

## Appendix A

The pretest and post-test questions are provided here.

*Students were given the following information:*
- The spin operators $\hat{S}_z$, $\hat{S}_x$, and $\hat{S}_y$ correspond to the observables $S_z$, $S_x$, and $S_y$, respectively, which in turn correspond to the z, x, and y-components of a particle's spin.
  - $\hat{S}_z|+z\rangle = \frac{\hbar}{2}|+z\rangle$ $\qquad\qquad \hat{S}_z|-z\rangle = -\frac{\hbar}{2}|-z\rangle$
  - $\hat{S}_x|+x\rangle = \frac{\hbar}{2}|+x\rangle$ $\qquad\qquad \hat{S}_x|-x\rangle = -\frac{\hbar}{2}|-x\rangle$
  - $\hat{S}_y|+y\rangle = \frac{\hbar}{2}|+y\rangle$ $\qquad\qquad \hat{S}_y|-y\rangle = -\frac{\hbar}{2}|-y\rangle$
- $|+z\rangle = \frac{1}{\sqrt{2}}|+x\rangle + \frac{1}{\sqrt{2}}|-x\rangle$ $\qquad\qquad |-z\rangle = \frac{1}{\sqrt{2}}|+x\rangle - \frac{1}{\sqrt{2}}|-x\rangle$
- $|+x\rangle = \frac{1}{\sqrt{2}}|+z\rangle + \frac{1}{\sqrt{2}}|-z\rangle$ $\qquad\qquad |-x\rangle = \frac{1}{\sqrt{2}}|+z\rangle - \frac{1}{\sqrt{2}}|-z\rangle$
- $a|+z\rangle + b|-z\rangle = \frac{a+b}{\sqrt{2}}|+x\rangle + \frac{a-b}{\sqrt{2}}|-x\rangle$
- $a|+x\rangle + b|-x\rangle = \frac{a+b}{\sqrt{2}}|+z\rangle + \frac{a-b}{\sqrt{2}}|-z\rangle$
- In all instances, $\hat{H} = C\hat{S}_z$, where C is a suitable constant. The energy eigenvalues are $E_+$ and $E_-$.

1. Consider a system in the state $|\chi\rangle = \frac{5}{13}|+z\rangle + \frac{12}{13}|-z\rangle$ with Hamiltonian $\hat{H} = C\hat{S}_z$. For simplicity, the energies corresponding to the $|+z\rangle$ and $|-z\rangle$ states are given as $E_+$ and $E_-$, respectively. Is the following statement true or false?

   $\hat{H}|\chi\rangle = E_+|+z\rangle$ or $E_-|-z\rangle$

2. Consider the normalized state $|\chi\rangle = a|+y\rangle + b|-y\rangle$. Are each of the following statements true or false?

   \_\_\_\_a) When an operator $\hat{S}_y$ acts on state $|\chi\rangle$, it is equivalent to the measurement of the observable $S_y$, and the measurement process is given by
   $$\hat{S}_y|\chi\rangle = \frac{\hbar}{2}|+y\rangle \text{ or } -\frac{\hbar}{2}|-y\rangle$$
   with $|\chi\rangle$ on the left-hand side representing the state before the measurement, and $|+y\rangle$ or $|-y\rangle$ on the right-hand side representing the state after the measurement.
   \_\_\_\_b) $|a|^2 + |b|^2 = 1$
   \_\_\_\_c) If $S_z$ (NOT $S_y$) is measured, the probability of obtaining $\frac{\hbar}{2}$ is $|a|^2$.

3. Consider a system in the state $|\chi\rangle = \left(\frac{2}{5} - \frac{2}{5}i\right)|+x\rangle + \frac{\sqrt{17}}{5}|-x\rangle$. If $S_x$ is measured, what are the possible outcomes, and what are their probabilities?

4. Consider a state $|\chi\rangle = \frac{4}{5}|+x\rangle + \frac{3}{5}|-x\rangle$ with Hamiltonian $\hat{H} = C\hat{S}_z$. (You may wish to refer to the information provided at the beginning of the pretest/post-test for some useful relations.)

a) If you measure $S_x$ and obtain a value $\frac{\hbar}{2}$, what is the normalized state of the system right after the measurement?

b) Immediately after you measure $S_x$ and obtain $\frac{\hbar}{2}$, you measure $S_x$ again. What is the probability of obtaining $-\frac{\hbar}{2}$?

c) Immediately after you measure $S_x$ and obtain $\frac{\hbar}{2}$, you measure $S_z$. What is the probability of obtaining $-\frac{\hbar}{2}$?

d) Immediately after you measure $S_x$ and obtain $\frac{\hbar}{2}$, you measure $S_z$ and then $S_x$ again, both in immediate succession. What is the probability of obtaining $-\frac{\hbar}{2}$ for this last measurement of $S_x$?

5. [Question asked in in-person class 2.]
Consider a system in the state $|\chi\rangle = \left(\frac{2}{5} - \frac{2}{5}i\right)|+x\rangle + \frac{\sqrt{17}}{5}|-x\rangle$. If many measurements of $S_x$ are made on identical systems prepared in this state, what is the expectation value of those measurements? Show or explain your work.

## Appendix B

The CQS questions are provided here. Correct answers are in bold type. As an additional note, the content that follows in these appendices is intended to help students understand the quantum measurement formalism, rather than technical details of quantum measurements in practice.

Notes to the instructor
- Students should be familiar with changing basis in two-state spin systems.
- CQS 1.1-1.3 revolve around the idea that measuring an observable in a given quantum state (a physical process involving an apparatus) is not the same as the corresponding operator acting on the state; specific cases are examined involving a spin operator, a Hamiltonian, and a generic operator $\hat{Q}$.
- CQS 2.1 helps students recognize that the expansion coefficients in a particular basis may be complex. CQS 2.2 emphasizes to students that the initial state must be written in the measurement basis associated with the observable to be able to determine the outcomes and the probabilities of measuring those outcomes. CQS 2.3 illustrates that, to reflect measurement collapse, the state after measurement must be normalized.
- CQS 3.1-3.2 focus on the collapse of a particular state into an eigenstate of the operator corresponding to the measured component of spin. They can help to reinforce that such eigenstates will result in 50/50 probabilities when either of the other two components of spin is subsequently and immediately measured.
- CQS 4.1-4.2 reinforce the concept of expectation value.

Notes to students
- All states appearing throughout are normalized, i.e., for a state $|\chi\rangle = a|\chi_1\rangle + b|\chi_2\rangle$, $|a|^2 + |b|^2 = 1$.
- The observable $S_i$ is the i-component of the spin. The corresponding operator is $\hat{S}_i$, for $i = \{x, y, z\}$.
- For instance, the observable $S_z$ is the z component of the spin, and the corresponding operator is $\hat{S}_z$.
- In all instances, $\hat{H} = C\hat{S}_z$, where C is a suitable constant.

Notes for CQS 1.1-1.2
- The following 2 questions present the same concept: once with a spin operator, and once with the Hamiltonian. They address what the application of a Hermitian operator is, and what it is not.
    o Application of an operator to a state *is not* the measurement of an observable.
    o Application of an operator to a state *is* a mathematical process that transforms the state.

- Option A, $\hat{H}|\chi\rangle = E|\chi\rangle$, for CQS 1.2 looks like the time-independent Schrödinger equation. However, $|\chi\rangle$ must be an eigenstate of the Hamiltonian for the equality to hold, known as "solving the energy eigenvalue problem."
- Note that students are assumed to know that if $\hat{H} \propto \hat{S}_z$, then $|E_\pm\rangle = |\pm z\rangle$.

## CQS 1.1
Which one of the following is correct if $|\chi\rangle = a|+z\rangle + b|-z\rangle$?
  A. $\hat{S}_z|\chi\rangle = \frac{\hbar}{2}|\chi\rangle$
  B. $\hat{S}_z|\chi\rangle = \frac{\hbar}{2}|+z\rangle$ or $-\frac{\hbar}{2}|-z\rangle$
  C. $\hat{S}_z|\chi\rangle = \frac{\hbar}{2}$ or $-\frac{\hbar}{2}$
  D. $\hat{S}_z|\chi\rangle = |+z\rangle$ or $|-z\rangle$
  E. **None of the above**

## CQS 1.2
Which one of the following is correct regarding the Hamiltonian operator $\hat{H}$ acting on a generic state $|\chi\rangle = a|+z\rangle + b|-z\rangle$?
(Note: Here, $\hat{H} = C\hat{S}_z$, where C is a suitable constant and, for simplicity, the energies corresponding to the $|+z\rangle$ and $|-z\rangle$ states are given as $E_+$ and $E_-$, respectively.)
  A. $\hat{H}|\chi\rangle = E|\chi\rangle$, where E is a constant
  B. $\hat{H}|\chi\rangle = E_+|+z\rangle$ or $E_-|-z\rangle$
  C. $\hat{H}|\chi\rangle = E_+$ or $E_-$
  D. **$\hat{H}|\chi\rangle = aE_+|+z\rangle + bE_-|-z\rangle$**
  E. None of the above

## CQS 1.3
Consider the following in the 2-D Hilbert space corresponding to electron spin:
- For every observable $Q$, there is a corresponding Hermitian operator $\hat{Q}$.
- The operator $\hat{Q}$ has two eigenstates, $|1\rangle$ and $|2\rangle$.
- The eigenstates are associated with the eigenvalues $q_1$ and $q_2$, such that $\hat{Q}|1\rangle = q_1|1\rangle$ and $\hat{Q}|2\rangle = q_2|2\rangle$.

Choose all of the statements that are correct about a measurement of the observable $Q$ made in the generic state $|\chi\rangle = a|1\rangle + b|2\rangle$.
1. The measurement of an observable $Q$ will collapse the state $|\chi\rangle$ into an eigenstate of the corresponding operator $\hat{Q}$.
2. A measurement of an observable $Q$ must return one of the eigenvalues $q$ of the Hermitian operator $\hat{Q}$.
3. An operator $\hat{Q}$ acting on state $|\chi\rangle$ is equivalent to the measurement of the observable $Q$. The measurement process is given by $\hat{Q}|\chi\rangle = q_1|1\rangle$ or $q_2|2\rangle$

A. 1 only  **B. 1 and 2 only**  C. 1 and 3 only   D. 2 and 3 only    E. All of the above

---

**CLASS DISCUSSION**

Notes for CQS 1.3

- The preceding question addresses the incorrect notion that making a measurement of an observable in a quantum state is equivalent to having the corresponding operator act on that state.
- In addition, other concepts related to measurement can be discussed:
    - Results of a measurement can only be eigenvalues of the operator corresponding to the observable being measured
    - Probabilities are determined by the state $|\chi\rangle$ at the time of measurement. The probabilities of obtaining $\lambda_i$ are governed by the Born rule: $|\langle \lambda_i|\chi\rangle|^2$ where $|\lambda_i\rangle$ is an eigenstate with eigenvalue $\lambda_i$ of the operator corresponding to the observable measured. The outcome of a measurement is, in general, not ensured, but can be predicted statistically.
    - The measurement collapses the state to the eigenstate associated with the eigenvalue that was measured.
- *Operators that commute:* When $\hat{H}$ and $\hat{S}_z$ commute, $[\hat{H},\hat{S}_z] = 0$ and the operators' normalized eigenvectors are identical. Eigenvalues are unique to each operator, with specified units ($\hat{H}$ has units of energy, and $\hat{S}_z$ has units of angular momentum).
- If Larmor precession has already been discussed, students can be reminded that the phenomenon is governed by the Hamiltonian (here $\hat{H} = C\hat{S}_z$, as is conventional, though of course it is equivalent if another component of spin is chosen to commute with the Hamiltonian, e.g., $\hat{H} = C\hat{S}_x$).

## CQS 2.1

Consider the state $|\chi\rangle = \frac{\sqrt{12}}{5}|+z\rangle + (\frac{3}{5} - \frac{2i}{5})|-z\rangle$. When a measurement of the z-component of the spin ($S_z$) is made in this state, which one of the following is true?

A. The probability of measuring $-\frac{\hbar}{2}$ is $(\frac{3}{5} - \frac{2i}{5})$.

**B. The probability of measuring $-\frac{\hbar}{2}$ is $\frac{13}{25}$.**

C. The probability of measuring $-\frac{\hbar}{2}$ is $\frac{9}{25}$.

D. The probability of measuring $-\frac{\hbar}{2}$ is $(\frac{3}{5} - \frac{2i}{5})^2$.

E. The probability of measuring $-\frac{\hbar}{2}$ is $\frac{1}{5}$.

Note for CQS 2.2: The following question may take students some time since it requires some calculation. Consider allowing 2-3 minutes.

## CQS 2.2

Consider the state $|\chi\rangle = \sqrt{\frac{3}{10}}|+z\rangle + \sqrt{\frac{7}{10}}|-z\rangle$. What is the probability of measuring a value of $+\frac{\hbar}{2}$ for the x-component of the spin ($S_x$)?

$$|+z\rangle = \frac{1}{\sqrt{2}}(|+x\rangle + |-x\rangle)$$
$$|-z\rangle = \frac{1}{\sqrt{2}}(|+x\rangle - |-x\rangle)$$

A) $\frac{3}{10}$

B) $\frac{(\sqrt{3}-\sqrt{7})^2}{20}$

C) $\frac{(\sqrt{3}+\sqrt{7})^2}{20}$

D) A measurement of the x−component of spin cannot be performed on a state $|\chi\rangle$ which is a superposition of eigenstates of $\hat{S}_z$.
E) None of the above

---

**CLASS DISCUSSION**
Notes for CQS 2.2
- At the end, tell students that it requires a change of basis, and ask some students how they did it.
- Go over the change of basis. This will help prime students for the following questions.

---

## CQS 2.3

Consider an electron spin in the state $|\chi\rangle = \sqrt{\frac{3}{10}}|+z\rangle + \sqrt{\frac{7}{10}}|-z\rangle$. The x component of its spin ($S_x$) was measured, and returned a value of $+\frac{\hbar}{2}$. What is the normalized state immediately after the measurement?

A) $\sqrt{\frac{3}{10}}|+x\rangle$
B) $\frac{\sqrt{3}+\sqrt{7}}{\sqrt{20}}|+x\rangle$
C) $\left(\frac{\sqrt{3}+\sqrt{7}}{\sqrt{20}}\right)^2|+x\rangle$
D) $|+x\rangle$
E) Not enough information

Note for CQS 3.1-3.2: The relationships between the $|\pm x\rangle$ and $|\pm y\rangle$ states are not given because it is not necessary for students to actually change to the $\{|+y\rangle, |-y\rangle\}$ basis and calculate the probabilities of measuring each outcome in order to correctly answer the questions.

## CQS 3.1

Consider the state $|\chi\rangle = \sqrt{\frac{2}{3}}|+x\rangle + \sqrt{\frac{1}{3}}|-x\rangle$. We first measure the observable $S_x$, then $S_y$ immediately after. Choose all of the following statements that are true:

I. For $S_x$, $P\left(\frac{\hbar}{2}\right) = \frac{2}{3}$, and $P\left(-\frac{\hbar}{2}\right) = \frac{1}{3}$.
II. For $S_y$, $P\left(\frac{\hbar}{2}\right) = \frac{2}{3}$, and $P\left(-\frac{\hbar}{2}\right) = \frac{1}{3}$.
III. For $S_y$, $P\left(\frac{\hbar}{2}\right) = \frac{1}{2}$, and $P\left(-\frac{\hbar}{2}\right) = \frac{1}{2}$.
IV. If $S_x$ was measured to be $\frac{\hbar}{2}$, then for $S_y$, $P\left(\frac{\hbar}{2}\right) = 1$; if $S_x$ was measured to be $-\frac{\hbar}{2}$, then for $S_y$, $P\left(-\frac{\hbar}{2}\right) = 1$.

A) I only
B) IV only
C) I and II only
**D) I and III only**
E) I and IV only

## CQS 3.2

Consider the state $|\chi\rangle = \sqrt{\frac{2}{3}}|+x\rangle + \sqrt{\frac{1}{3}}|-x\rangle$. We first measure the observable $S_y$, then $S_x$ immediately after. Choose all of the following statements that are true:
   I. For $S_y$, $P\left(\frac{\hbar}{2}\right) = \frac{2}{3}$, and $P\left(-\frac{\hbar}{2}\right) = \frac{1}{3}$.
   II. For $S_x$, $P\left(\frac{\hbar}{2}\right) = \frac{2}{3}$, and $P\left(-\frac{\hbar}{2}\right) = \frac{1}{3}$.
   III. For $S_x$, $P\left(\frac{\hbar}{2}\right) = \frac{1}{2}$, and $P\left(-\frac{\hbar}{2}\right) = \frac{1}{2}$.
   IV. If $S_y$ was measured to be $\frac{\hbar}{2}$, then for $S_x$, $P\left(\frac{\hbar}{2}\right) = 1$; if $S_y$ was measured to be $-\frac{\hbar}{2}$, then for $S_x$, $P\left(-\frac{\hbar}{2}\right) = 1$.
   A. I only
   B. II only
   **C. III only**
   D. II and IV
   E. III and IV

### CLASS DISCUSSION

Notes for CQS 3.1-3.2
- Note that in the preceding questions CQS 3.1-3.2, option IV can never be true because $\hat{S}_x$ and $\hat{S}_y$ do not commute.
- CQS 3.1 invites a discussion that, once the state has collapsed to an eigenstate of one of the spin operators, measurement (in immediate succession) of the same spin component does not change the state, while measurement of either of the other two spin components will return spin-up and spin-down with equal probability.

## CQS 4.1
Choose all of the correct expressions for the expectation value $\langle S_y \rangle$ in a state $|\chi\rangle = a|+y\rangle + b|-y\rangle$.
   I. $\langle \chi|\hat{S}_y|\chi\rangle$
   II. $|a|^2\langle+y|\hat{S}_y|+y\rangle + |b|^2\langle-y|\hat{S}_y|-y\rangle$
   III. $(|a|^2 - |b|^2)\frac{\hbar}{2}$
   A. I only
   B. III only
   C. I and II only
   D. I and III only
   **E. All of the above**

## CQS 4.2
Given a state $|\chi\rangle = \frac{3}{5}|+y\rangle + \frac{4}{5}|-y\rangle$, which of the following is the expectation value $\langle S_y \rangle$?
   A. $-\frac{7}{50}\hbar$
   B. $-\frac{1}{10}\hbar$

C. Either $+\frac{\hbar}{2}$ or $-\frac{\hbar}{2}$
D. 0
E. None of the above

**CLASS DISCUSSION**

Notes for CQS 4.1-4.2
- Expectation values can be introduced in several complementary ways:
    - Using Dirac notation $\langle\chi|\hat{Q}|\chi\rangle$
    - Using matrix representation in a given basis to compute $\langle\chi|\hat{Q}|\chi\rangle$
    - Characterizing expectation value as a weighted average of measurement outcomes on an ensemble of identically prepared systems, i.e., $\sum_i$ (probability of measuring ith eigenvalue) × (value of ith eigenvalue)
    - Showing the equivalence of all these approaches